# Effect of surface morphology and magnetic impurities on the electronic structure in cobalt-doped BaFe$_2$As$_2$ superconductors

Qiang Zou[1,*], Zhiming Wu[1,2,*], Mingming Fu[2], Chunmiao Zhang[2], Shivani Rajput[1,3], Yaping Wu[2], Li Li[3], David S. Parker[3], Junyong Kang[2], Athena S. Sefat[3], Zheng Gai[1,&]

[1] Center for Nanophase Materials Sciences, Oak Ridge National Laboratory, Oak Ridge, TN 37831, USA

[2] Fujian Provincial Key Laboratory of Semiconductors and Applications, Collaborative Innovation Center for Optoelectronic Semiconductors and Efficient Devices, Department of Physics, Xiamen University, Xiamen, Fujian Province, 361005, P. R. China

[3] Materials Science & Technology Division, Oak Ridge National Laboratory, Oak Ridge, TN 37831, USA

[*] These authors contributed equally to this manuscript
[&] Corresponding author: *gaiz@ornl.gov*

Combined scanning tunneling microscopy, spectroscopy and local barrier height (LBH) studies show that low-temperature-cleaved optimally-doped Ba(Fe$_{1-x}$Co$_x$)$_2$As$_2$ crystals with x=0.06, with $T_c$ = 22 K, have complicated morphologies. Although the cleavage surface and hence the morphologies are variable, the superconducting gap maps show the same gap widths and nanometer size inhomogeneities irrelevant to the morphology. Based on the spectroscopy and LBH maps, the bright patches and dark stripes in the morphologies are identified as Ba and As dominated surface terminations, respectively. Magnetic impurities, possibly due to cobalt or Fe atoms, are believed to create local in-gap state and in addition suppress the superconducting coherence peaks. This study will clarify the confusion on the cleavage surface terminations of the Fe-based superconductors, and its relation with the electronic structures.

One major difference between the iron-based superconductors (FeSC) and conventional superconductors is the necessary involvement of magnetic elements. For a long time magnetic elements had been thought to suppress the superconductivity because of the competitive nature of superconductivity and magnetism, but with the cuprates and now FeSC, there are two cases of high-temperature superconductivity that apparently are magnetic in origin [1-4].

FeSC share a common Fe$_2$X$_2$ layer structure where X is a pnictogen (P, As) or a chalcogen (S, Se, Te). With the incorporation of intercalation layers between the Fe$_2$X$_2$ layers, FeSC are classified into five families: '11', '111', '122', '1111' and complex structure materials [3-4]. To understand the causes of this superconductivity, the relationship between the global properties and local behavior must be understood. These local signatures include chemical defects, electronic variations, dopant distributions, and strain. The '11' and '111' families have been studied extensively using scanning tunneling microscopy and spectroscopy (STM/STS) and the results are clear due to the cleanness of the cleavage as there is unique and symmetric cleavage plane, and a non-polar surface with minimal surface effect [5-6]. On the contrary, the STM/STS studies on the '122' family are still controversial although great efforts have been put in the studies [5-12], with main debate focused on where and how the crystal cleaves. For example, within the BaFe$_2$As$_2$ (Ba122) unit cell, the Fe-As layers have a mirror symmetry around the Ba layer as shown in Fig. 1 (a). As FeAs layer is a highly polar plane, the physical cleavage through this plane might force Ba ions to be divided to both sides of the cleavage surface to alleviate the polar nature, associated with surface reconstructing. As a result, rich morphologies or "terminations" exist on the Ba122 cleavage surface. Some groups report that the crystal cleaves at Ba plane with ½ monolayer of Ba atoms to achieve the charge neutral surface[7-10], while others suggest As atoms terminate the surface by breaking the weak bond between As and Ba [11-12].

In this work on optimally-doped Ba(Fe$_{1-x}$Co$_x$)$_2$As$_2$ crystal (Co-Ba122), we find that different "terminations" or morphologies can be observed even from the same batch of crystals and same low-temperature cleavage condition, indicating the similar surface free energy of the different configurations. On the other hand, the superconducting gap maps from different terminations are not affected by surface terminations, owing to the global nature of the superconductivity. The Ba or As dominated surfaces can be directly identified by combining the STM, STS and local barrier height (LBH) information. In addition, non-superconducting impurities are observed in the superconducting map. We suspect the magnetic impurities are possibly cobalt or Fe atoms, showing an in-gap state and suppressing the coherence peaks of the superconductivity.

Single crystals of Ba(Fe$_{1-x}$Co$_x$)$_2$As$_2$ were grown out of self-flux, and FeAs and CoAs binaries synthesized similar to our previous reports [13]. Phase purity, crystallinity, and the atomic occupancy of crystals were checked by collecting Powder X-ray diffraction (XRD) data on an X'Pert PRO MPD diffractometer (Cu $K_{\alpha 1}$ radiation, $\lambda$=1.540598 Å); The average chemical composition of each crystal was measured with a Hitachi S3400 scanning electron microscope operating at 20 kV, and use of energy-dispersive x-ray spectroscopy (EDS).

The single crystals of Co-Ba122 (x = 0.06 dopant levels) were cleaved in ultra-high vacuum (UHV) at ~ 100 K and then immediately transferred to the STM head which was precooled to 4.2 K. The STM/STS experiments were carried out at 4.2 K in UHV low-temperature scanning tunneling microscope with base pressure of $2\times10^{-10}$ Torr, with mechanically cut Pt-Ir tip. All Pt-Ir tips were conditioned and checked using clean Au (111) surface before each measurement. Topographic images were acquired in constant current mode with the bias voltage applied to the samples. All the spectroscopies were obtained using the lock-in technique with bias modulation $V_{rms}$ = 0.5 mV at 973 Hz. Spectroscopic imaging was preformed over a grid of points (64 × 64 pixels) at various energies using the same lock-in amplifier parameters[14-15]. The local barrier height (LBH) mapping was carried out over a grid of points (64 × 64 pixels) at set point $Z_0$ (I = 100 pA, V = 50 mV), the current I (z) was measured as the tip retracting from $Z_0$ to $Z_0$ + 100 pm with feedback loop off, $\Delta Z_{rms}$ = 10 pm [16-17]. The local barrier height is the square of slope of *lnI vs z* at each pixel.

The surface topographic images of Co-Ba122 are shown in Figure 1. Fig. 1(b) shows a cleaved surface morphology of the crystal. The height of the steps is 2 nm as shown in the line profile Fig. 1(b), 1.5 times of the c lattice constant of Co-Ba122 reflecting its mirror symmetry within the primitive unit cell [18]. From this image and the line profile, it is obvious there are corrugations rather than atomically flat surfaces. It is still a controversy about the cleaved surface morphology in atomic level of Ba122, depending on the cleavage condition or crystals [6, 11]. On our samples, after extensive survey on different crystals but same cleaving conditions, we summarize the typical images in Fig. 1(c)-(f). In brief, the majority areas in Fig. 1(c) are 2 × 1 stripe structure similar to the one reported in literature[8]; in Fig. 1(d) and (e), the stripes are gradually separated by brighter area; in Fig. 1(f), the brighter areas are fully percolated, with meandered defect boundary lines as reported [7, 12]. Atomic resolution images of 2 × 1 stripe and √2×√2 reconstructions can be achieved from Fig. 1(c) and (f), as shown in the insets, with reconstructions outlined with red and green arrays, respectively. Fig. 1(g) displays an area with coexistence of the 2 × 1 stripe domain (domain 1) and √2×√2 area (domain 2). As shown in the line profile of Fig. 1(g), there are angstrom-sized height variations on the terrace, similar on both domains (note the different Z scale in Fig. 1(b) and (g)).

The area with the coexistence of the two characteristic domains provides the best opportunity to explore the correlations between the different surface morphologies with the properties of the crystals on the Co-Ba122, as the STM morphology is the constant local density state of the surface which relies on their environment [6]. The current-imaging-tunneling spectroscopies (CITS) were carried out on domain 1 and 2. Fig. 2(a) to (d) show the topographies of domain 1 and 2 and their corresponding normalized dI/dV spectroscopies averaged over the whole area, respectively. Although the morphologies are totally different, the two normalized dI/dV spectra surprisingly show very similar line shape. The normalized tunneling current dI/dV(v) can be fitted by using Dynes model[19] as

$$\frac{dI(V)}{dV} = \int_{-\infty}^{\infty} dE \, \text{Re} \left\{ \frac{E - i\Gamma}{\sqrt{(E - i\Gamma)^2 - \Delta^2}} \right\} \left\{ \frac{1}{k_B T} \frac{(E + eV/k_B T)}{[1 + \exp(E + eV/k_B T)]^2} \right\},$$

where $\Delta$ is the superconducting gap function, T is the temperature, $\Gamma$ is the broadening parameter. S-wave gap function was used to fit the normalized dI/dV. During the fitting, energy dependent broadening parameter $\Gamma$ ($\Gamma(E) = \Gamma_1 + \Gamma_2 *E$) were used [20]. The fitting curve are overlaid with the normalized dI/dV for both domains in Fig. 2(b) and (d), respectively, with $\Delta$ values of 3.96 meV and 4.08 meV. We also tried the d-wave gap fitting; it is interesting to note that our data does not show any preference for s-wave or d-wave gap function.

The data of Figures 2(b) and 2(d) displays a notable lack of strong coherence peaks, as being typically observed in canonical s-wave superconductors such as Pb. This is often attributed to, and modeled by, strong scattering (i.e. a large Dynes $\Gamma$). However, such a large and energy-independent $\Gamma$ leads to a zero-bias conductance far in excess of the values observed in Figures 2(b) and (d), and a *d*-wave expression does not yield a significantly better fit. Given this difficult situation, we recalled the work of Alldredge et al [20], in which cuprate dI/dV curves were fit using a $\Gamma$ linear in energy – i.e. $\Gamma(E) = \alpha *E$, with α dependent on location. Experimentally speaking, this form allowed the successful fitting in that work of dI/dV spectra with neither substantial coherence peaks nor significant zero-bias conductance. Such a $\Gamma(E)$ form is believed related to the nodal nature of superconductivity in the cuprates. In fact, using this form directly does not fit these data properly, but a related form $\Gamma(E) = \Gamma_1 + \Gamma_2 *E$ in fact does fit the data rather well. The implications of this fit are somewhat difficult to understand. The energy-linear term could plausibly relate to a nodal superconductivity, with the energy-independent term describing scattering due to impurities or other defects. In this vein one recalls that these materials [21] are known theoretically to be subject to a competition between the $s_{+/-}$ state proposed by Mazin et al [22] and a *d*-wave state, with the pairing strength of the two very nearly equal. What state is ultimately selected depends on parameters such as the Hund's rule J which presumably varies with material. In this context the combination of energy-*independent* and energy-*linear* terms in Γ(E) is reminiscent of this competition, though it would require much more involved theoretical work to explain this fully.

To reveal the intrinsic differences from the two domains, we look into the spatial dependent superconductivity gap maps derived from the pixel by pixel dI/dV spectra fitting of those two domains with the s-wave Dynes and Γ(E) model. Fig 2(e) shows the superconducting gap map simultaneously collected with Fig. 2(c) in domain 2. The gap map shows nanometer size inhomogeneity as widely reported in literature for these Fe-based superconductors [8, 23]. There also exist non-superconducting locations in the map, similar to the reports on the same material [24]. The gap map from domain 1 is basically the same as the domain 2 map. Fig. 2(f) shows the statistical distributions of the gap Δ from the two gap maps. The center of the distributions of superconducting gaps is at 4.5 meV.

The similarity of the superconducting behavior from different morphologies is reasonable since the superconductivity is a bulk global property rather than a local behavior, considering the 3 nm coherence length of pairing electrons found for Co-Ba122 [8]. The influences of the rich surface morphologies should be reflected in the local density of state (LDOS) as it is more termination, composition and reconstruction related. Figure 3(a) show the comparison of the average LDOS (dI/dV(I/V)) from the two domains. Although inside the gap they are similar, the out-of-gap features from the two domains are very different, due to the contribution from the non-superconducting states and temperature broadening effect. Fig. 3(b) and (c) are the pair of STM image and LDOS slice at -6.5 mV from domain 1. The bright and dark features in the Fig. 3(b) are well correlated to the morphological features in Fig. 3(b). The average LDOS from the bright and dark areas in domain 1 are plotted in Fig. 3 (d), the domain 1 and 2 data are also reproduced in this figure for comparison. The fact that the LDOS from the bright area of domain 1 overlaps with the average data from whole domain 2 strongly suggests that domain 2 are dominant by the bright areas of domain 1, which is consistent with our observation of the gradually developed morphology in Fig. 1.

To identify the surface chemistry of the domains, the barrier height mapping was used to measure the local barrier height (work function) of the material [16-17]. In principle, the tunneling current *I(s)* exponentially decays with the tip-sample distance z, $I(z) \propto \exp(-\sqrt{8\Phi m_e/\hbar^2}\, z)$, $\Phi = 0.952 * (\frac{dlnI}{dz})^2$, where Φ is the tunneling barrier height, approximately equal to the average of the tip and sample work function[16]. Fig. 4 (a) and (b) are the pair of STM image and the spatial dependent LBH image from domain 1. The spatial information of barrier height is just the reminiscences of the topography of the surface. The bright and dark areas in the STM image related to the cooler and warmer color patches in the LBH image, meaning lower or higher relative work functions, respectively. The top of Fig. 4(c) displays a line profile, in terms of height variation of the surface, acquired along the marked white dash line of Fig. 4(a). It has about 0.2 nm height difference between the bright patches and the dark stripes. The Ba-122 crystal has layered structure, with Fe-As tetrahedral layer as the basic building block and Ba layers sandwiched in between the Fe-As layer as shown in the bottom of Fig. 4(c). The Fe cations are combined with As anions as covalent-dominated bonds, while both Ba-As and Ba-Ba bond are ionic-dominated bonding. During cleavage, Ba-122 tends to break with Fe-As layer intact, but with Ba atoms left on both side of the cleaved surface because of ionic bonding between Ba-As and Ba-Ba, from the Fig. 4(c) model, the height between the Ba and As layer is about 0.2 nm.

By combining all the experiment results above, we believe the bright patches are Ba dominated while the dark stripes are As dominated. The 0.2 nm height difference between the bright patch and the dark stripes in the Fig.4(c) consistent with the height model; the cooler and warmer colors in the local barrier height image of Fig. 4(b) reflect the lower or higher work functions of Ba and As, as reported [17, 25], correspondingly the bright patches and dark stripes in morphology; The LDOS below the Fermi level in Fig. 3(d) from the bright patches are smaller than that from the dark stripes, consistent with the less density of states from 0.5 monolayer Ba-termination surface comparing to that of the As-termination, from the first-principle calculation of the Ba-122 surface [26]; The atomically resolved $\sqrt{2}\times\sqrt{2}$ reconstructions from the bright patches and the $2 \times 1$ stripe from the dark areas in Fig. 1(c)-(f) also consistent with the claims from literature reports, as Ba or As dominated reconstructions [6]. In brief, the domain 1 and domain 2 discussed in this paper are As and Ba dominated termination, respectively. We note here that the domain size of the bright patches and the stripes are few nanometers, smaller than the coherence length of the probing electrons of low energy electron diffraction (LEED), we anticipant the features are difficult to be caught by LEED IV studies because of the scattering from the small domains [11].

With the above knowledge of the surface chemistry of the complicated morphology of the Co-Ba122, we can analyze more details from the images to explore the understanding of the material. As shown in Fig. 3(d), although Ba and As areas show large differences below the Fermi level, they are quite similar above Fermi level right above the SC gap. We anticipant that imaging at those energy range would exclude the Ba and As involved effect and extract information about the dopants and impurities. Showing in Fig. 5(a) and (b) are a pair of STM and dI/dV maps (+20 mV) from an area mainly composed of As dominated reconstructions with some Ba patches. As expected, the majority areas of the Fig. 5(b) do not have much contrast, except the red circled areas. Those red circled protrusions exist in LDOS maps even down to Fermi level. Those areas are also circled red in the STM image in Fig. 5(a) correspondingly, they are protrusions higher than the Ba patches. Actually there are more protrusions in the STM image but no contrast in the dI/dV map, some of them were circled in blue.

The dI/dV spectroscopies along the blue arrow in Fig. 5(c) which crosses one of the red circled protrusion was displayed in Fig. 5(d). In these set of dI/dV spectra, the superconducting gap gradually broaden when approaching to the protrusion. In the meantime, about nanometer away from the morphological protrusion, an in-gap peak appears close to the Fermi level in the spectra within the gap, and the peak intensity enhances gradually when approaching to the protrusion. Contour map of the spectra in Fig. 5(e) clearly shows the development of the impurity state circled in red in the gap. The results indicate that the red circled protrusions behave as pair breaker to suppress the coherence peaks locally. It is difficult to identify the chemical origin of the protrusions using only our data. As reported, nonmagnetic impurities, for example Cu and S, have no influences on the coherence peaks[27-28], while magnetic impurities like Mn[28] and Fe[29] will create new in gap state and suppress the superconducting gap feature. We assume the impurities observed here are of magnetic nature based on the comparison that the visible in gap state is similar to the magnetic impurities feature but different from the non-magnetic ones. Considering the elements in our crystals, we suspect those impurities might be Co or Fe atoms. Those impurities are not the Co dopants that replace the Fe atoms in lattice to create the superconductivity of the crystal, the pairing type of the crystal should not be changed by those small amounts of impurities.

We comprehensively studied the low-temperature cleaved surface of optimal doped Ba(Fe$_{1-x}$Co$_x$)$_2$As$_2$ single crystals with $T_c$ = 22 K using low temperature STM. Various surface morphologies with bright or dark areas coexist in our cleavage conditions, but the superconducting gap maps show same gap width and similar nanometer size inhomogeneity, owing to the global nature of the superconductivity. But the variation in the morphologies shows vast differences in the LDOS and LBH maps. The bright patches and dark stripes were identified as Ba or As dominated surface terminations. Locally appeared in-gap state as well as the suppressed coherence peaks of the SC is observed around some atomic level protrusions, which we suspect as magnetic impurities. In summary, combined STM, STS and LBH studies show that no matter how complicated the morphologies are, the surfaces are the mixture of the Ba dominated bright patches and As dominated stripes. We believe this study clarifies the confusion on the terminations of the cleavage surface of the Fe-based 122 superconductors.

**Acknowledgements**

The research is equally supported by the U.S. Department of Energy (DOE), Office of Science, Basic Energy Sciences (BES), the Center for Nanophase Materials Sciences, and Materials Science and Engineering Division (MSD). STM/S experiments were conducted at the Center for Nanophase Materials Sciences, which is a DOE Office of Science User Facility. Z. Wu team's work was supported by the National Key Research and Development Program of China (No. 2016YFB0400801) and the National Natural Science Foundations of China (no. 61227009).

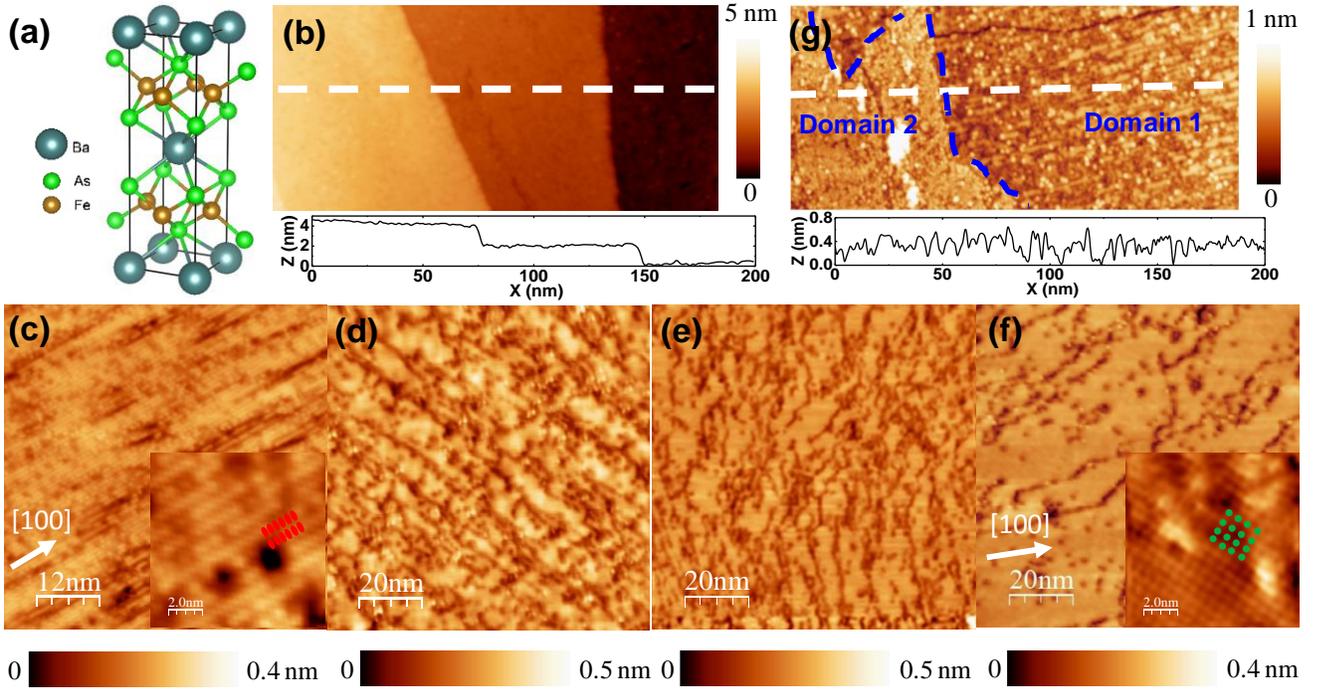

Figure 1 **STM Morphologies of low-temperature-cleaved optimal doped Ba(Fe$_{1-x}$Co$_x$)$_2$As$_2$ single crystals.** (a) Crystal structure of Ba122. (b) A typical STM morphology of Co-Ba122 with bias voltage of V$_{bias}$= 100 mV and tunneling current of I$_t$ = 50 pA. The profile of the white line is shown at the bottom of Fig. (b). (c)-(f) Summary of characterized STM morphologies from various cleavages and locations, with V$_{bias}$= 20 mV and I$_t$ = 50 pA. The insets of Fig. 1(c) and (f) are the atomic resolution images of 2 × 1 stripes and √2×√2 reconstructions, respectively. (g) STM morphology with two domains coexist, V$_{bias}$= 20 mV and I$_t$ = 100 pA. The blue dash line separates the two domains, in which domain 1 shows 2 × 1 strip feature, and domain 2 shows √2×√2 feature. The bottom of Fig. (g) shows the line profile along the white line.

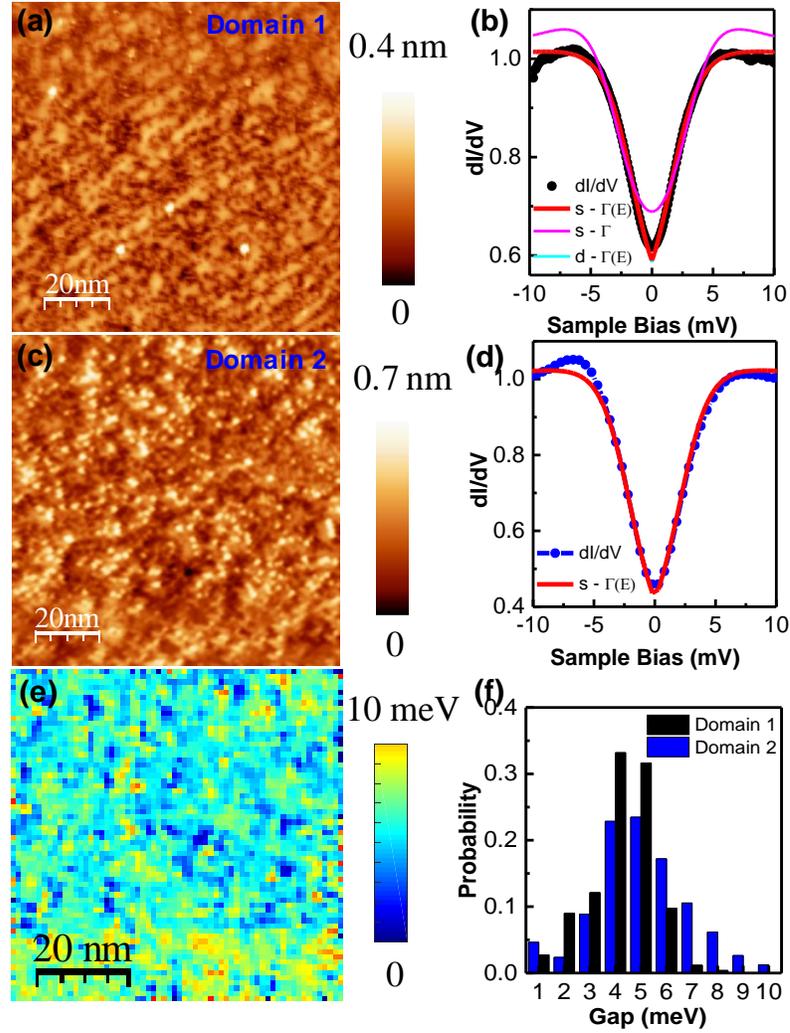

Figure 2 **Similar superconductor gaps from two domains.** (a) and (b) Morphology and average dI/dV spectroscopy of domain 1 with $V_{bias}$= 10 mV and $I_t$ = 100 pA. (c) and (d) Morphology and average dI/dV spectroscopy of domain 2 with $V_{bias}$= 20 mV and $I_t$ = 200 pA. The red line is the Dynes s-wave $\Gamma(E)$ fitting of the dI/dV spectra in (b) and (d). The non-energy dependent s-wave $\Gamma$ and d-wave $\Gamma(E)$ fitting are also shown in (b). (e) The SC gap map simultaneously collected with (c) in domain 2. (f) The statistic distributions of the SC gap $\Delta$ from two domains.

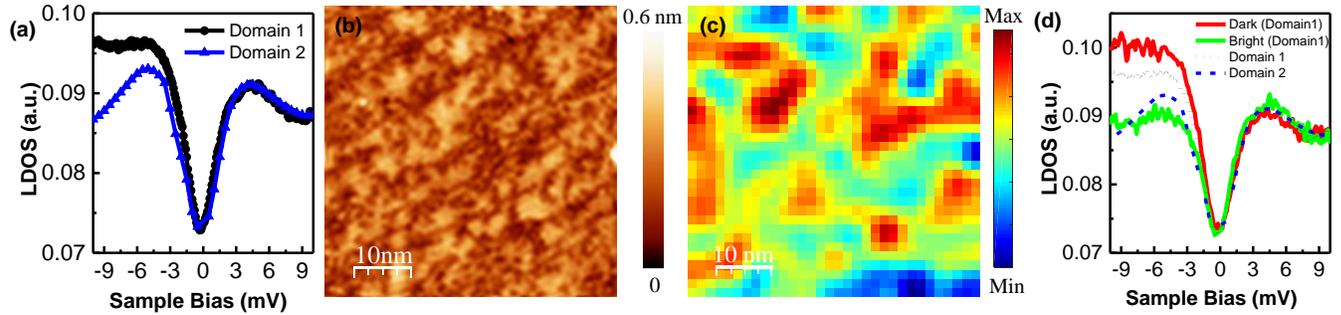

Figure 3 **Comparison of local density of states from different domains.** (a) Averaged local density of states of domain 1 and 2 from Fig. 2(a) and (c). (b) and (c) A pair of STM morphology and LDOS slice (at -6.5 mV) of domain 1, $V_{bias}$= 20 mV and $I_t$ = 100 pA. (d) Averaged LDOS spectra from the bright and dark areas of (b), the spectra from whole domain 1 and 2 are also plotted for comparison.

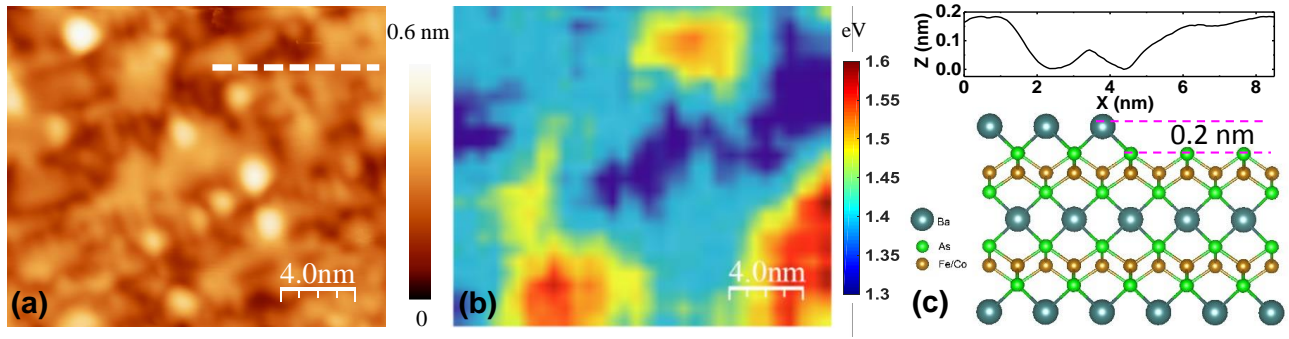

Figure 4 **Surface morphology and local barrier height image (LBH) identify surface layers**. Morphology (a) and simultaneously acquired LBH map (b) acquired from domain 1, $V_{bias}$= 50 mV, $I_t$ = 100 pA. The warmer color in (b) represents a higher local barrier height. (c) Top: the profile of the white line in (a); Bottom: the proposed model of surface terminations in domain 1.

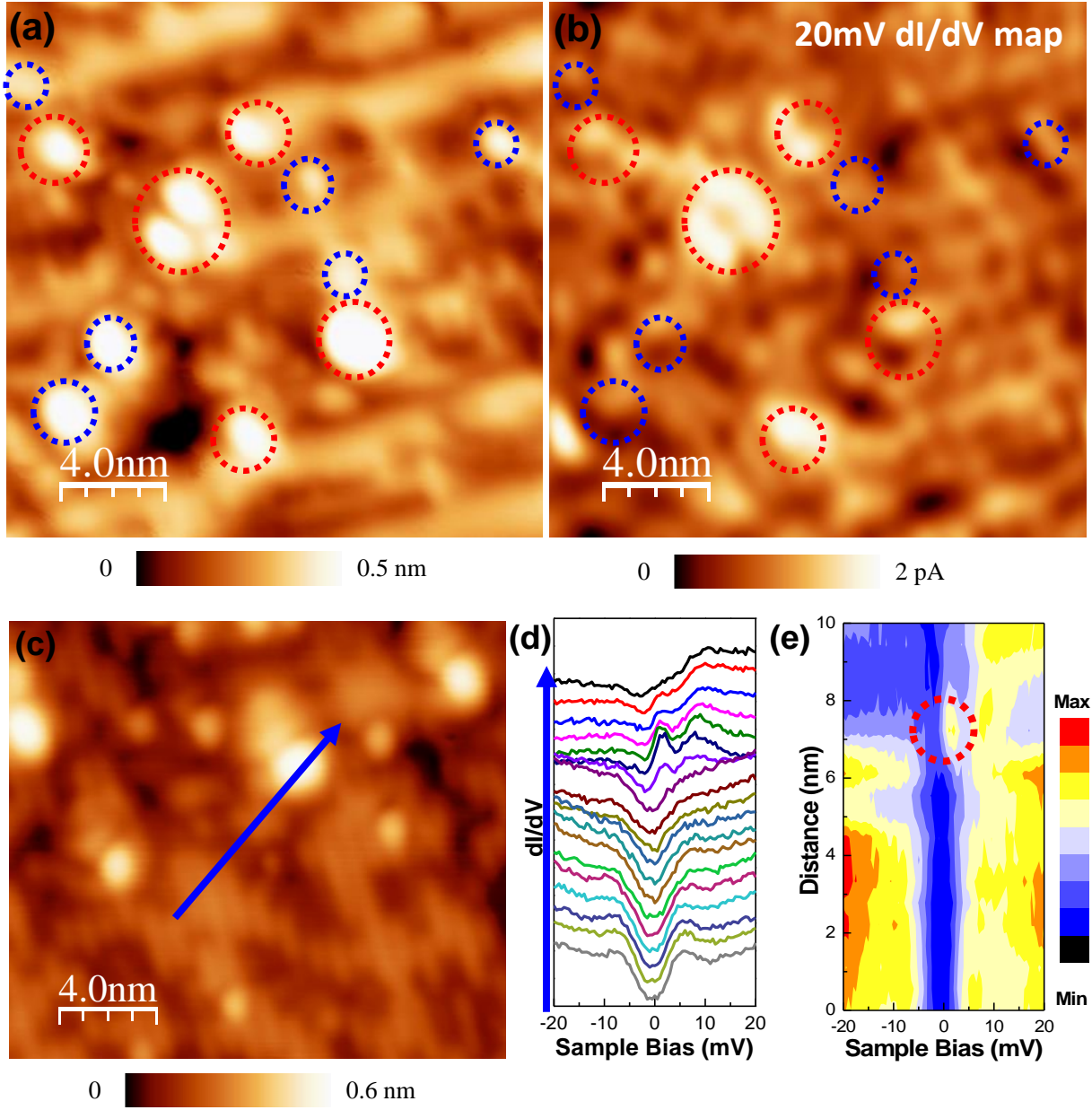

Figure 5 **Impurity states suppress the coherence peak of SC.** (a) and (b) A pair of STM and dI/dV map from domain 1, with $V_{bias}$= 20 mV and $I_t$ = 100 pA. The red circled protrusions exist high LDOS while the blue has no contrast on LDOS at 20 mV. (c) - (e) Spatial evolution of STS along the blue arrow in (c). The spectroscopies in (d) are offset for clarity. The SC coherence peaks gradually broaden when approaching to the protrusion. An in-gap peak appears close to the Fermi level about nanometer away from the protrusion as shown in the red circle in (e).